# Cross-Platform DNA Methylation Classifier for the Eight Molecular Subtypes of Group 3 & 4 Medulloblastoma

Omer Abid, Gholamreza Rafiee*

*Abstract*— Medulloblastoma is a malignant pediatric brain cancer, and the discovery of molecular subgroups is enabling personalized treatment strategies. In 2019, a consensus identified eight novel subtypes within Groups 3 and 4, each displaying heterogeneous characteristics. Classifiers are essential for translating these findings into clinical practice by supporting clinical trials, personalized therapy development and application, and patient monitoring. This study presents a DNA methylation-based, cross-platform machine learning classifier capable of distinguishing these subtypes on both HM450 and EPIC methylation array samples. Across two independent test sets, the model achieved weighted F1 = 0.95 and balanced accuracy = 0.957, consistent across platforms. As the first cross-platform solution, it provides backward compatibility while extending applicability to a newer platform, also enhancing accessibility. It also has the potential to become the first publicly available classifier for these subtypes once deployed through a web application, as planned in the future. This work overall takes steps in the direction of advancing precision medicine and improving clinical outcomes for patients within the majority prevalence medulloblastoma subgroups, groups 3 and 4.

## I. INTRODUCTION

Medulloblastoma is a malignant brain cancer widely known for its prevalence in children. Through extensive treatment strategies based on surgery, chemotherapy and radiation therapy, approximately 75% of the patient are able to survive in the long term [1]. These treatments while crucial also come along with negative side effects, effecting patients' lives [1] [2], especially considering the implications on the growing children. However, with advancement in genomics, molecular subgroups have been discovered within the disease. These subgroups have shown to be heterogenous in clinical, biological and outcomes perspective [3]. These in fact are now considered better definition of disease behaviour than conventional techniques [3]. This has led to a push towards personalizing treatments and directed therapies focusing on the goals of achieving better outcomes and quality of life for patients.

In 2012, the international pediatric oncology community agreed on the presence of four medulloblastoma subgroups: WNT, SHH, Group 3 & Group 4, displaying differing genetic features, clinical features and prognosis [4]. During this conference, the community also acknowledged the presence of subtypes under majority of these subgroups but they were not well described. [4]. In 2016, after much research published in regards to these subgroups, the world health organization recognized 4 medulloblastoma subgroups (WNT, SHH-TP53 - Wild Type, SHH-TP53-Mutant, Non -WNT/ Non - SHH) [5], with the last category hosting both group 3 and 4.

The group 3 and group 4 subgroups are particularly important entities as they constitute a majority 65% of all medulloblastoma cases and had shown differing outcomes, linkage with high risk disease factor and considerable relapses despite the absence of the risk factors [6]. Three studies in 2017 suggested the presence of 8 [7], 4 [8] and 6 [9] subtypes under group 3 & group 4 using DNA methylation data and also integrating gene expression data for the latter study. In 2019, a study, aimed to harmonized the results of these studies by applying the methodologies used in each of these studies to a combined larger cohort created from these studies along with consensus analysis and biological information analysis. This resulted in the identification of consensus 8 subtypes (I - VIII) showing differing age, cytogenetic characteristics and survival profiles [6]. Fast forward in 2021, these 8 subtype have been recognized in the latest 5th edition of WHO classification of tumors of the central nervous system [10].

The above advancement in understanding the disease landscape is crucial for building personalized treatments [6], however, the ability to efficiently, reliably and accurately classify subgroups on patient samples is as important, as classification is important to the development of personalized therapies [3], aid in clinical trials [11], and application of personalized therapies and monitoring standard [12] once these have been developed. In turn in a more abstract manner this means improving outcomes for patients and also reducing therapy related negative side effects due to the treatment strategies based on previous knowledge. However, no publicly available classification solution is yet known to integrate the 8 newly identified subtypes under group 3 and 4.

It is important to note that the identification of the medulloblastoma subgroups in the above studies have been genomic data-driven in nature focusing on the analysis of gene expression and DNA methylation data in cohorts and the methods used within these are not capable of single sample classification. Genome wide DNA methylation is understood as a gold standard in classifying medulloblastoma subgroups [11]. It is the addition of methyl within genomic areas which influences the expression of genes and deviations within it, are known to be associated with development of diseases and cancers [13]. Currently the Illumina Infinium HumanMethylation assays are considered to be the most cost-effective platform for DNA methylation profiling [14]. Over time Illumina has introduced multiple assays such as HM450,

Omer Abid and Gholamreza Rafiee are with the School of Electronics, Electrical Engineering and Computer Science (EEECS) at Queen's University Belfast (QUB), Email: {oabid01, G.Rafiee}@qub.ac.uk

EPIC and EPICv2, with each measuring methylation at 450 000, 850,000 and 950 000 genomes sites (also called CPG sites/ CPG Probe) respectively. Apart from latter, all have been deprecated. This assay method does not directly provide interpretable data to judge methylation and data generated from these assays is required to be fed to specific software which then provide quality check tools, processing tools and finally generate beta values which are interpretable DNA methylation values against each CPG probes for each sample. The beta values are ranged in between 0 to 1, where 0 denote no methylation of molecules and 1 denotes majority methylation of molecules.

Given the data orientated nature, machine learning (ML) based classification tools have been popular solutions as its methods are primarily focused on learning from data and dealing with high dimensional data. These are able to predict single samples and are accurate, efficient and reliable in predictions. Many effective and accurate ML based classification tools have been built since the identification of the four consensus subgroups [11][12][15] and as new subgroups have been discovered updated tools have been built [16] and some solutions have even explored MRI scans as the data platform [17], essentially experimenting with a data platform other than the original driving platforms.

Based on the above and the established significance of group 3 & 4 patients along with the newly identified heterogenous 8 subtypes under this category, this study aims to explore an machine learning based classification solution for these newly identified 8 subtypes focusing on accuracy and reliability. Such a tool would aid in the development of personalized therapies, aid in clinical trials and application of personalized therapies and monitoring standard once these are developed. Additionally, during building of such tools further important links between the subtypes and the predicting factors could be discovered highlighting new knowledge which could in turn benefit in building of personalized treatment. This study covers the end to end process of model building including analytics and robust evaluation. The methodology focuses on explaining the procedures and methodologies used for dataset search, dataset preparation, data preprocessing, pre and post model analytics and model training, tuning and evaluation. In the sections below first a literature review is presented investigating availability of ML based classifiers for different subgroups, the data platforms used and their effectiveness, and the pre-processing steps and AI algorithms used. Next, the methodologies are presented followed by results, discussions and conclusion.

## II. LITERATURE REVIEW

To build an effective solution, it is crucial to gain insights from existing ML solutions. Hence, this section at first investigates availability of ML classifiers for subgroups to justify need. Next it presents data platforms utilized and their success, and finally it analyses the preprocessing and algorithms used. This review is limited to exploring ML focused studies.

### A. Medulloblastoma Subgroups & Availability of Classifiers

Considering the 4 subgroups identified in 2012, in the period from 2018 to 2025, 4 classifiers have been published catering to these subgroups. [11][12][15][17] with 3 of them published in 2023 and onwards. Similarly considering the advancements in subgroups as noted in 2017 and 2019, one classifier was found to be published in 2025 [16], catering to 7 subgroups, incorporating two new subtypes under each of group 3 & 4 [8] ,and one classifier was built within the original 8 subtypes – group 3 and group 4 consensus study for these subtypes, however, the classifier is not publicly available [6]. This indicates: first, the solution efforts seem to be focused on the four subgroups as models are still being continuously built for these, Second, the incorporation of new subgroups after their discovery in 2017 and consensus in 2019 has fairly lagged as only two solutions catering to any of these have been published until now, with one not present in the public domain. While the improvement aspect in relation to the 4 consensus subgroups is great, it is important to focus on incorporating new knowledge and ensuring its public availability especially when it has been officially accepted by WHO, such as the case of eight new subtypes of group 3 and group 4 [10]. Lastly, additional search suggested that a solution for these subtypes is not publicly available in non-ML based solutions as well.

### B. Data Platforms Used & Their Effectiveness

Considering the same classifiers, 4 of these were built utilizing HM450 DNA methylation beta values [6][11][12][16]. Two of these, dealt with four subgroups, one using a two model classification approach having an a lowest accuracy 99.4% and 92.2% across different validation sets for the three and two subgroup models respectively [11], and other using a single model, having an average accuracy of 96% [12]. The third model had a mean balanced accuracy of 98% for 7 subgroups [16]. The fourth model considering the 8 subtypes for group 3 and 4 had a AUC of 0.9969 [6]. Similarly one model used gene expression data to predict four consensus subgroups with the accuracy of above 90% across various conditions [15]. Lastly, another model used a combination of computer and human derived features from MRI images to predict the four consensus subgroups in a two model approach achieving an average micro-accuracy of 76% and 91% respectively, across different validations [17]. These performances indicate that DNA methylation and gene expression based classifiers have comparable performances, however, the two methylation classifier dealing with 7 subgroup and 8 subtypes, with excellent performance, establishes DNA methylation platform's ability in segregating more granular level of medulloblastoma subtypes. The radiomics based classifier although having a much cheaper assay method (MRI scans) and also having the benefit of being non-invasive in nature compared to the other two data platforms lacked in accuracy for the 4 subgroups. Keeping this in mind, the classification task for the 8 new subtypes is more complex, and the accuracy is of immense importance in medical domain. Additionally, none of the DNA methylation classifiers were built of on newer version of methylation arrays: EPIC and EPICv2, while the higher resolution can improve accuracy and the ability to classify increasing subgroup heterogeneity. Moreover, a cross platform classifier have never been attempted. Given the probe content overlap across the array versions, such a solution could effectively predict for sample assayed on the newer version of array by training on the prior array version when direct training on newer version is not possible due to sample scarcity. It also provides accessibility of the classifier to samples on different array version and backward compatibility.

*C. Preprocessing & Machine Learning Algorithms*

Across the studies on methylation data [6][11][12][16], the 8-subtype study selected the top 50,000 and top 10,000 highest standard deviation (SD) probes and performed feature selection on it [6]. The 7-subgroup classifier study selected top 10,000 most variable probes and applied non-negative matrix factorization (NMF) to summarize probe content and reduce the data to 6 dimensions [16]. Among the 4 subgroups classifiers, the two-model study selected 5904 and 2612 most variable probes using SD for each of the model and then applied custom feature selection, identifying differentiated probes for each subgroup to reduce the dimensions [11]. The single layer 4 subgroup model selected the top 5000 most variable cpg probes using mean absolute deviation and it used a random forest-based feature selection [12]. All the studies except from the 8-subtype study [6] used more than one model types [11][12][16] and each had different best performing models. Comparing these studies the major difference lies in feature selection approach and models used. The approach of selecting more probe content and then using NMF to summarize probes using meta-genes seems to be the better approach than selecting the most relevant probes (selecting a very limited number) as it can even utilize those probes which are relevant but left out due to the dimension reduction task (essentially by feature selection) by other methods. In addition, selecting medium to high correlated probes to the subgroups before NMF, may help to reduce data noise as selecting only the most variable probes may contain non-correlating probes adding noise.

*D. Conclusion*

To conclude, it is evident that a publicly available classification solution for 8 subtypes under group 3 and 4 is missing. A DNA methylation-based solution seems to be the most suitable given its ability to segregate increasing heterogeneity and accuracy. From dataset search it is found that no samples exist on EPICv2 and scarcely exist on EPIC. Hence, a cross-platform classifier, trained on HM450 data while accommodating the most feasible version from recent array version (EPIC) is the best solution while also providing backwards compatibility with HM450. Lastly, an approach based on selecting a higher number of most variable probes, possibly 25000, and then selecting medium to highly correlated probes before performing NMF to summarize the probe content could possibly reduce data noise and improve results.

III. METHODS

This section lists the methodologies for the cross platform solution. For details on the alternative EPIC only classifier solution please refer to the supporting material.

*A. Dataset Search*

A dataset of 1501 samples was identified directly from the Group 3/4 - 8 subtypes consensus study [6]. This dataset was present under the series GSE130051 on GEO – Gene expression omnibus, a public repository which hosts genomic datasets from research studies. This series contained raw IDAT files for 1501 samples which contained 1391 HM450 and 110 EPIC assayed samples and the processed HM450 based beta values for all 1501 samples (containing 453152 CPG probs), along with miniml file which hosted the metadata for these 1501 samples. A further dataset search activity on GEO was performed, however, no other datasets having the annotations for these 8 novel subtypes were found against HM450, EPIC and EPICv2.

*B. Dataset Preparation*

Due to the presence of raw IDAT files (intermediate files produced by DNA methylation arrays assay), EPIC beta values for 110 samples out of the total 1501 samples could be generated. This provided an opportunity to generate a primary dataset of processed HM450 beta values keeping only those samples which were assayed on HM450 for model training and internal validation, and then have two platform specific validation sets consisting of samples for which HM450 beta values were already available and also EPIC beta values could be generated through DNA methylation processing library. This allowed for a robust cross platform solution building and evaluation opportunity. First to generate the EPIC validation set, the miniml file from the GEO was parsed through methylprep 1.7.1 to generate sample metadata files against each array version. Using methylcheck 0.8.5, the metadata for EPIC samples was read in and based on that IDAT files for 110 EPIC assayed samples were identified. These were then processed in R with SeSAMe 1.26 [18], being recommend as one of the third party tools to process DNA methylation data by Illumina for its arrays [19]. First basic quality control check was performed (checking for overall detection success rate, checking dye bias, signal background, and bisulfite conversion success on a randomly chosen sample). Next, signal intensities data was generated using sesame's recommended processing pipeline for EPIC samples. This pipeline included applying quality Mask (used to mask study independent poor design probes based on the recommended mask identified for EPIC array from research [20], see Table I for details on what types of probes are masked and reason for it), inferInfiniumIChannel (infer the channel), dyeBiasNL (dye bias non-linear correction) pOOBAH (p-value based masking with standard SeSAMe threshold of 0.05), noob (normalization). The processed signal intensities were re-analyzed with the same QC on the same randomly selected samples to check for improvements. No problematic issues were noted and the beta values were generated. The beta values were loaded in python and samples which did not have the subgroup identified were filtered resulting in an EPIC validation set of 99 samples and 866553 CPG Probes. Next, HM450 processed beta values for 1501 samples were loaded. Due to its bulky nature (12 GB in size – 1501 samples with 453152 columns), the loading was chunked and filtered for unreliable probes based on the quality mask for HM450 (retrieved from SeSAMe based on the library's established credibility, see Table I for further details on filtration criteria and number of probes filtered), probes on X,Y chromosomes to remove sex bias (X,Y chromosome are sex linked [21]) and non-cg probes as medulloblastoma studies have focused on cg probes only. The list of probes on X and Y chromosomes was also retrieved from SeSAMe based on the latest genome build for HM450. Additionally, sample filtration was also performed at load time to remove samples which did not have any subtype annotated. This resulted in a dataset of 1370 samples and 381810 probes. This data was then divided into a primary set containing samples only assayed on the HM450 platform (1271 samples) and a HM450 beta values validation set containing 99 samples which were essentially the same samples as in the EPIC validation set. Finally, the primary set was split into test and train set using a stratified 80-20 split resulting in a train set of 1016 samples and internal test set of 255 samples. The stratification was applied due to indication of class imbalance, and the split was performed even before any of data preprocessing/pre-model analytics to avoid any data leak within the internal test to simulate a real validation scenario even with the internal test set. See Table II for details on different sets.

TABLE I. PROBES FILTERED UNDER THE RECOMMENDED QUALITY MASK USED IN SESAME BASED OFF RESEACH [20]. SESAME PERFORMS THE SAME MASKING FOR HM450 AS WELL

| Filter Name | Category |
|---|---|
| Mapping - Probes with mapping issues | polymorphism |
| channel_switch – channel switching probes | polymorphism |
| snp5_GMAF1p - SNPs using global allele frequency of 1% with 5 base pair | polymorphism |
| extension | polymorphism |
| sub30_copy | Non-uniqueness |
| EPIC = 105454 Probes<br>HM450 = 64144 Probes | |

## C. Data Preprocessing

The train set was checked for duplicate samples based on GEO GSM Ids (This are unique Ids given to each sample present on GEO), and beta value distribution (see supporting material Figure 1) and beta values ranges were checked. However, no duplicate samples, problematic beta value distribution or ranges were noted. Missing value analysis was performed and it indicated no missing values. Next, the top 25 000 probes were selected by ranking the standard distribution of each CPG probe in descending order. The logic behind this was that methylation/de-methylation events are linked to disease development and characteristics, hence the above selected top 25000 such probes with high variability in methylation. Next these probes were further filtered using the criteria of having an absolute correlation of greater than 0.4 with the subtypes, resulting into the selection of 13931 probes. The intent was to reduce noise from data by removing probes with very less predictive value and keeping mildly and highly correlated probes so that these could be incorporated within NMF summarization to find latent features. Next, for a cross platform solution to work it was necessary that from these highly correlated probes only those were selected that were also present in the EPIC assay so that NMF metagenes could be projected from train set to the test sets. It is important to perform the NMF projection as performing NMF independently tends to find metagenes based on the samples present within the set and could differ although not vastly but enough to cause the model to fail on the test sets. It was also noted that common probes in between EPIC and HM450 are directly comparable [22]. Hence there was no concern about differing beta values for the same probes in between the two platforms for the same samples. Based on this notion a list of reliable EPIC probes was curated by removing probes based the same methodology and logic as done during the loading of the HM450 processed beta values for 1501 samples. This time, however, the quality mask (see Table I for details on the types of probes filtered and the total number) and X, Y chromosomes probes list retrieved from SeSAMe were for the EPIC platform. The X, Y probes list was according to the latest genome build. After removal of these along with cg probes, the resulting reliable probes were 741145. An intersection of the this with the highly correlated probes were performed resulting in 13916 probes. See supporting material Figure 2 for correlation heatmap with subtypes based on these final selected probes. A final data level duplicate check was also performed, however no duplicates were found. NMF was selected as the dimension reduction technique to further reduce the feature space fit for model building based on the literature review and given NMF's success in identifying metagenes against medulloblastoma subtypes [16]. However, since the number of components (metagenes) K and NMF's solver and beta loss parameters were treated as hyperparameter, hence this was performed during the model hyper parameter tuning step. Next, for the internal test and both the validation sets, for all, beta value distributions (see supporting material Figure 3,4,5) and beta values ranges were checked to ensure that these were logically valid datasets, however, no problematic behaviour was noted. 13916 probes from the train set were directly selected for each of these sets. Next missing values analysis was performed on all the sets, and only the EPIC validation set was found to have missing values. Samples with more than 30% of missing probe values were dropped based on the logic than with such a degree of missingness even with imputation, these samples would not be reliable. This resulted in 2 samples to be dropped from this set. For imputing the remaining missing values, two strategies were used, first, for probes for which no values were present at all in the EPIC validation set, the values from the HM450 validation set were imputed against each corresponding sample. This was logical same probes in between these two platforms are directly comparable [22]. Second, for probes with intermittent missing values, KNN imputation with 5 neighbors was performed. The reason for choosing KNN imputation was that same medulloblastoma subgroups samples show similar DNA methylation patterns as indicated by studies [11][12], and KNN works by finding similar samples and then imputes missingness using data from these similar samples. It is important to note that KNN imputation is performed after selecting the probes which help to avoid unnecessary imputation and reduce dimensionality for the KNN imputation task. Lastly the imputation was performed with a weight based on distance to have more influence of most similar samples and the selection of the number of neighbors parameter was done based on visualizing random CPG probes' imputed values and overall distribution pattern (see supporting material Figure 6).

TABLE II. DESCRIPTION OF TRAIN SET, INTERNEL TEST SET, HM450 AND EPIC VALIDATION SETS. NO OTHER INFORMATION APART FROM THE SUBTYPES WAS KNOWN ABOUT THE SAMPLES

| Set Name | Number Of Samples Against Subtypes | | | | | | | | Total Sample |
|---|---|---|---|---|---|---|---|---|---|
| | 1 | 2 | 3 | 4 | 5 | 6 | 7 | 8 | |
| Train Set | 33 | 130 | 93 | 105 | 81 | 93 | 224 | 257 | 1016 |
| Internal Test | 8 | 33 | 23 | 26 | 20 | 23 | 57 | 65 | 255 |
| HM450 Validation | 10 | 14 | 8 | 9 | 7 | 4 | 22 | 25 | 99 |
| EPIC Validation | 10 | 14 | 7 | 8 | 7 | 4 | 22 | 25 | 97* |

\* 2 sample dropped due to more than 30% missingness during validation set preparation at end of preprocessing.

### D. Pre-Model & Post Model Analytics

Pre-model analytics consisted of visualizing beta value density plot, class imbalance using a count plot, and 3-D principal component analysis (PCA) scatter plot and t-distributed Stochastic Neighbor Embedding (TSNE) based scatter plot to visualize the grouping of samples in relation to the subtypes in the reduced dimension spaces. Both PCA and TSNE are unsupervised dimension reduction techniques. PCA is based explaining variability in data while TSNE has been popular with respect to genomic data analysis. For PCA, the first 3 principal components were considered and for TSNE the first 100 principal components from PCA were used to generate a reduced 2 dimensional representation of the data (It is recommend to reduce the dimensions first by PCA and then apply TSNE). Additionally for TSNE, parameters such as perplexity, learning rate and early exaggeration were identified based on error and trial. (See supporting for further pre-model analytics performed). It is important to note that all the pre-model analytics were only performed on the train set to avoid any data leak. Post model, the same PCA, TSNE, NMF metagenes heatmap visualizations were used for internal test set. NMF metagenes heatmap with best K was performed for train set as well to analyze the pattern of metagenes against subtypes.

### E. Model Training, Hyper Parameter Tuning and Evaluation

Four model were selected to be trained. SVC was chosen based on its ability to classify complex non-linear and linear problems. Decision Tree was chosen for its ability to work on dataset where cutoff points could stratify the target classes and random forest was chosen as an ensemble version of it as ensemble models are robust to overfitting and have greater predictive power based on combining multiple tree models. Last but not the least KNN was chosen based on the knowledge of the dataset that same subgroups have similar methylation pattern for their samples [11][12], this should also emit in NMF metagenes and based on this similar subtype samples should cluster in space, on which KNN could work very well. This approach reflected a notion of a greedy approach by choosing models of different capability on the dataset and is common in research studies. The main strategy was to treat NMF K, solver and beta_loss as hyper parameters and for each NMF K, solver and beta loss combination perform a 10-fold stratified (class imbalance was noted in train set) cross validated hyper-parameter tuning using gridsearch for all the models on the train set, and test the tuned models on the internal test set. This would result in tuned models' cross validation and test scores across varying K, solver and beta loss combinations based on which the best model, best K, beta_loss and solver could be selected. The reason to use such a nested approach is that when hyperparameter tuning and several model are evaluated in one go reporting just the cross validation score for selecting amongst the models constitutes a data leak for each model this score was optimized in finding the best hyperparameter (a train aspect). Within this approach NMF reduction was applied on the entire cross validation train set first. For cross validation an imblearn pipeline was created for each model to apply minmax scaling (as NMF metagenes had varying scales), random under sampling and over sampling to train folds and then project minmax scaling on test folds to minimize any form of data leak in cross validation. The under sampling and oversampling was performed due class imbalance in train set, the samples for subtype 7 and 8 were very high as compared to other subtype samples so these were under sampled. The subtype 1 had lower samples as compared to all the other class despite under sampling so all the classes except the majority class were oversampled to reach a balance. It is important to note that the oversampling was performed with a shrinkage of 0.1 to add some random small variability to the newly added samples instead of bootstrapping on the notion of emulating a real scenario as same subtype sample are close but vary. Based on the internal test score, decision of choosing the model amongst the 4 models was made and then the best K, NMF solver and beta_loss were selected based on the hyper-parameter tuning cross validation score for this model. This model with the best hyperparameters for selected NMF parameters was then evaluated on internal test, HM450 validation set and EPIC validation set. The prior was considered as a base model and hence an in-depth second phase of 10-fold stratified hyperparameter tuning with a refined grid was also performed using gridsearch for further model hyper-parameter tuning on the selected NMF parameters. This was then re-evaluated extensively on internal test, HM450 and EPIC validation sets. In terms of hyperparameter grid, the first hyper-parameter tuning cycle considered choosing a wide search approach for each model selecting a wide range of values for each parameter and then for in-depth hyperparameter tuning for the best model the grid was modified to focus on areas around the best model hyper-parameters found in the initial hyperparameter tuning for this model. Please see Table III and for further details please see supporting material.

## IV. RESULTS

### A. Software And Hardware Enviornment

Software wise this study used SeSAMe package 1.26 within R 4.5 for IDAT file processing to generate beta values for EPIC validation set while all the rest of the work was performed within a Jupyter notebook using python. Activities of dataset preparation, pre-processing and pre-model analytics were performed on Kaggle using the latest python version, due to the RAM intensive nature of dataset preparation and pre-processing.

TABLE III. HYPER PARAMETER GRID FOR EACH MODEL IN INTIAL TUNING AND REFINED GRID FOR BEST MODEL IN DEPTH TUNING

| Model | Model Hyper-Parameter Grid |
|---|---|
| SVM | C: [0.05,1,4,7]   kernel: 'poly', 'rbf', 'sigmoid'<br>degree: [1,3,5]   gamma: [0.0,1,3,5,7] |
| KNN | n_neighbors: [3,6,9,12,24,36]   weights: ('uniform', 'distance') |
| DT | criterion: 'gini', 'entropy', 'log_loss'   max_depth : [3,6,9,12]<br>ccp_alpha: [0.01, 0.1,1,5] |
| RF | Same as DT |
| Model | Refined Grid for Best Model |
|  | C: 0.01 to 0.95 with increments of 0.05   kernel: 'poly'<br>degree: [2,3,4]   gamma: 1.1 to 4.85 with increments of 0.05 |

Model training/tuning, evaluation and post-modal analytics were performed on personal machine due to the compute intensive nature of model training/tuning tasks, using Jupyter notebook from the Anaconda distribution with python version 3.12.7. In terms of hardware, on Kaggle, a 4 core CPU with 30GB RAM was utilized on the platform (less than 30GB RAM is not sufficient for the activities of dataset preparation and pre-processing). For personal machine, a 6-core, core i7, 8$^{th}$ generation CPU with 16GB RAM was utilized. For further in-depth details please refer to supporting material.

### B. Pre-model Analytics

Overall pre-model analytics indicated a presence of class imbalance as indicated by the count plot in Fig 1. Subtypes 7 and 8 had a high occurrence while subtype 1 had the lowest occurrence, and the rest of the subtypes were almost equivalent. Moreover, both the 3-D PCA and TSNE visualizations (Fig 1) showed that same subtype samples clustered in the reduced dimensional space indicating the similarity of methylation characteristics of samples in each subtype. For these it is also important to note that subtype clusters are always neighbored with other subtype clusters having the closest subtype number (e.g. 2 neighbored 1, 3), show casing a relation.

### C. Selection of evaluation metric

In terms of evaluation metrics, during any hyperparameter tuning tasks, F1-Macro average score was chosen as the metric to be optimized. This metric incorporate precision and recall (is the harmonic mean of these), and calculates it for each target class and macro-averages it. This is suitable to assess the average correct identification power of each subtype and this is crucial as each subtype is equally important. The reason for choosing 'Macro' was to give equal importance to each subtype's performance regardless of the number of samples it contained making the optimization goal to find the best identification ability for all subgroups regardless of their occurrence. This is more robust to balanced accuracy as it only incorporate recall as the measure of positive identification power, while F1 score incorporates recall and precision to access the positive identification power. For evaluation of model on test/validation sets, the strategy was to perform an extensive evaluation, so balanced accuracy, weighted f1-score, confusion matrix to analyze classification errors, classification report to gain in depth details on precision of each subtype were used to judge about the identification power for each subtype. Weighted F1-score was selected to be the final reported metric of classifier accuracy so that class imbalance in test/validation sets did not penalize the score unnecessarily and instead the true capability of correctly identifying each subtype is presented. For the final tuned model's evaluation on test/validation sets, in addition to all the above, incorrect prediction analysis with prediction probabilities to judge about incorrect predictions, and threshold-precision/recall tradeoff for each subtype were also analyzed to further evaluate probability based prediction perspective of the model. ROC – curves with area under the curve (AUC) was not performed due its biasness when the positive class examples are less, as this was the case in the evaluation of each subtype in an one-vs-all analysis. Precision-recall curve, which focus on positive class and do not suffer from the above biasness were used with AUC for each subtype to judge model's robustness.

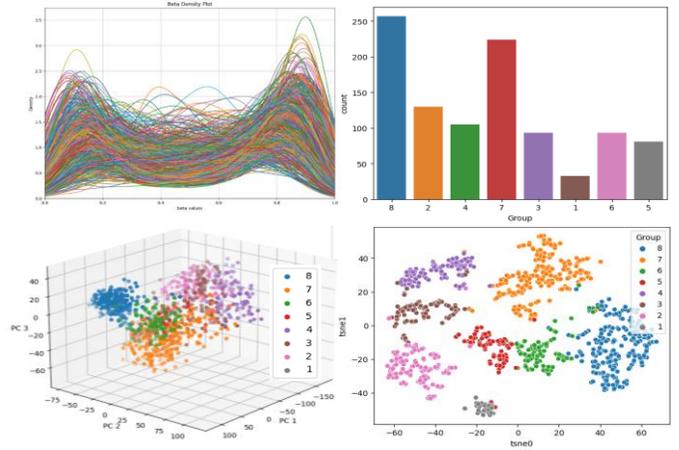

Fig. 1. Collage of pre-modal analytics performed on train set. **Top left:** Beta value distribution using 13916 probes. **Top right:** Count plot of subtypes in train set. **Bottom left:** 3-D PCA scatter plot with subtypes, **Bottom right:** TSNE scatter plot with subtypes.

### D. Model Results

The nested hyper-parameter tuning activity showed that KNN and SVC had superior performance on internal test set compared to all other models regardless of the choice of NMF K and other NMF parameters. In between these two, SVC in general, performed better on most K and other NMF parameter combinations, hence SVC was selected as the best model (see Fig 2 left section). For SVC, the cross-validation hypermeter tuning scores indicated the best performance at NMF K = 17 using multiplicative as solver and kullback-leibler as beta loss with the F1-Macro score of 0.974 (see Fig 2 right section), and the best hyper-parameters for SVC were noted to be: C: 0.05, degree: 3, gamma: 3 and kernel: poly. The main evaluation metric scores for this model are presented in Table IV for the 3 test sets, however, for classification reports please refer to supporting materials Figure 7, 8 and 9. This model achieved an average weighted F1-Score of 0.947 across the test sets, demonstrating a generous ability to correctly identify each subtype. It is important to note that the weighted F1-score on EPIC validation set was slightly lower than the other sets (0.94 as compared to 0.95 for the other two). However, the model still performed very well on this set, demonstrating its ability to perform predictions on EPIC assayed samples. Looking at the confusion Matrices for the 3 test sets (Fig 3), the model mostly correctly identified all the subtypes indicated by the diagonal pattern in the matrices. All 3 test sets displayed similar patterns of miss-classification with miss-classification usually happening with neighboring subtypes. The confusion matrix for EPIC and HM450 sets were also very similar, demonstrating that the model performed similarly on both the platforms. With in-depth hypermeter-tuning, the following model hyper-parameters were found to be the best one: C: 0.06 (stored as $6.0000000000000005 \times 10^{-2}$), degree :3, gamma: 3.15 (stored as 3.1500000000000017) and kernel: poly. (Please see supporting materials Figure 10 for further details on in-depth hyper-parameter tuning). For this model, the weighted F1-Score remained the same for HM450 set and internal test set, however, it improved a bit on the EPIC validation set, improving the overall average weighted F1-score to 0.95. This indicated an increase in the model's ability to perform on EPIC assayed samples. See Table V for main evaluation metrics for this model and supporting materials Figure 11, 12 and 13 for classification reports for each test set. The confusion matrix remained the same for HM450 set, improved a bit for EPIC validation set and for internal test set one additional sample was miss-classified, when compared with the base tuned model matrices (see Fig 4 for confusion matrices). Overall, the analysis remained the same as presented for the base tuned model confusion matrices. For all 3 test sets, the analysis of incorrect predictions (predicted versus true subtypes) showed that all incorrect predictions happened with neighboring subtypes shown in the TSNE scatter plot in Fig 1. Analysis in terms of probability-based prediction, showed that for internal test set, for 11 out of 13 incorrect predictions, the second highest probability belonged to the correct subtype. For the other two sets, all the incorrect predictions had the second highest probability belonging to the correct subtype. This indicated that in probability-based predictions this model had a very robust second guess. (Please see supporting materials Tables 1-3 for incorrect predicted versus true subtypes along with prediction probabilities for all 3 test sets.) Precision-recall curves for each subtype, except subtype 3 and 4, across all test sets almost perfectly followed the right upper corner indicating a robust model working at all probability thresholds. Subtype 3 and 4's curves were slightly deviated showing a similar pattern across all validation sets. The average area under the precision recall curve is presented in Table V and across all validation sets, the average area is 0.953 which indicates a good robust model. Please see supporting material Figure 14, 15 & 16 for all the precision-recall curves across all test sets. An analysis of threshold and precision-recall trade-off was also conducted across all validation sets and for each subtype. However, every subtype showed a different best threshold so a universal probability-based prediction threshold could not be derived. Please see supporting material figures 17-19 for threshold and precision-recall trade-off curves for all test sets.

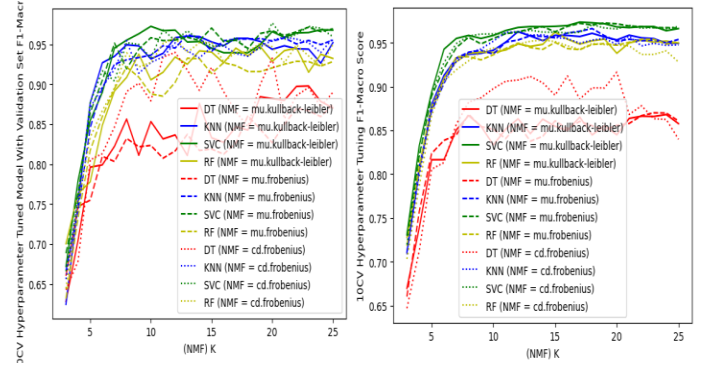

Fig. 2. **Left**: Internal test set F1-Macro average score of the 4 models presented across varying NMF K and other NMF parameters. The color of the line denotes the model type (i.e. SVC, KNN) and line style represents different solver and beta loss combination. **Right**: Same as Left, however with train set hyper-parameter tuning cross validation F1-Macro average score.

TABLE IV. MAIN EVALUATION SCORES FOR SVC INITIAL TUNED MODEL AND IN DEPTH TUNED MODEL

| Evaluation Statistics For tuned base model | Set Name | | | |
|---|---|---|---|---|
| Scoring (Rounded to 2-decimal place for individual set scores and 3 for overall score) | Internal Test Set (n = 255) | HM450 Validation set (n = 99) | EPIC Validation Set (n = 97) | Average Scores across all sets |
| Macro F1-Score | 0.94 | 0.94 | 0.91 | 0.930 |
| Weighted F1-Score | 0.95 | 0.95 | 0.94 | 0.947 |
| Balanced Accuracy | 0.94 | 0.97 | 0.94 | 0.950 |

TABLE V. MAIN EVALUATION SCORES FOR SVC IN DEPTH TUNED MODEL.

| Evaluation Statistics For in depth tuned final model | Set Name | | | |
|---|---|---|---|---|
| Scoring (Rounded to 2-decimal place for individual set scores and 3 for overall score) | Internal Test Set (n = 255) | HM450 Validation set (n = 99) | EPIC Validation Set (n = 97) | Average Scores across all sets |
| Macro F1-Score | 0.94 | 0.94 | 0.93 | 0.937 |
| Weighted F1-Score | 0.95 | 0.95 | 0.95 | 0.950 |
| Balanced Accuracy | 0.94 | 0.97 | 0.96 | 0.957 |
| Average Precision-Recall AUC | 0.96 | 0.95 | 0.95 | 0.953 |

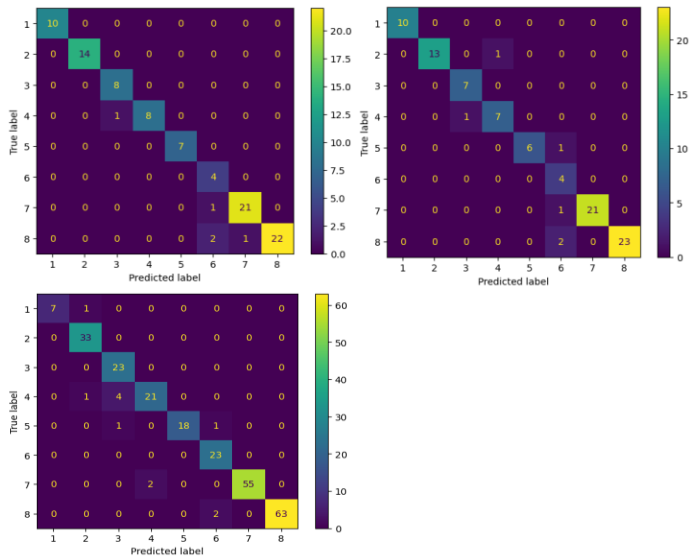

Fig. 3. Confusion matrices for base tuned model. **Top Left**: HM450 validation set **Top Right**: EPIC validation set **Bottom**: Internal Test set. Labels are subtype identifiers (1-8)

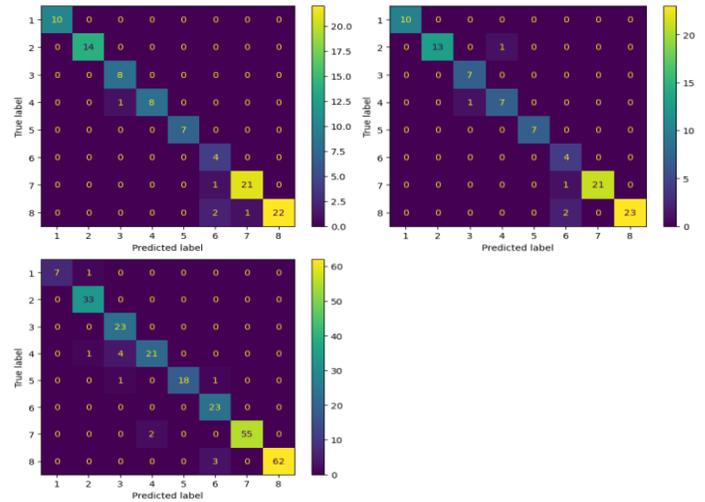

Fig. 4. Confusion matrices for in depth tuned model **Top Left**: HM450 validation set **Top Right**: EPIC validation set **Bottom**: Internal Test set. Labels are subtype identifiers (1-8)

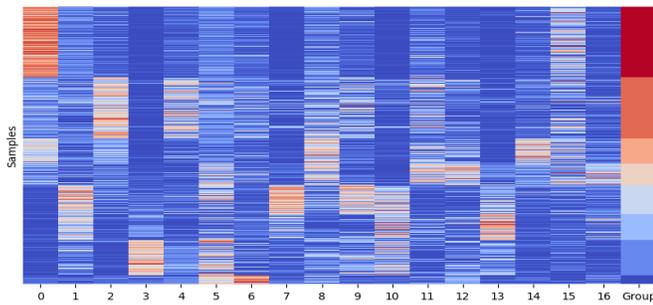

Fig. 5. 17 metagenes (min-max transformed) heatmap visualization of samples sorted based on subtype. Dark blue contour in group column represents subtype 1 and it increases to red as moving towards subtype 8. It is important to note that the metagenes show distinguishing patterns for each subtype.

To conclude, the final model proved to be robust and accurate. It performed equally well on samples from both HM450 and EPIC platforms, establishing success as a cross-platform solution. Lastly it is important to note that only comparison of this classifier could be performed with the classifier presented in the original 8 subtypes under group 3 and 4 consensus study. The study indicated an area under curve score of 0.9969 [6] which is presumed to be under the ROC curve. A direct comparison cannot be made due to the discussed biasness of ROC curves and hence these were not generated, but the model built is comparable and has the additional benefit of being cross platform in nature, hence valuably adding to the research.

## V. Discussions, AI Ethics and Future Work

This study built an accurate and reliable cross platform classification solution for the purpose of classifying 8 novel medulloblastoma subtypes under group 3 and group 4, achieving an average weighted F1-score of 0.95 and average balanced accuracy of 0.957 across 3 test sets. While this classifier could not be publicly deployed due to the time constraint of this study, however, it did complete the first step in making the first publicly available classification solution for these subtypes, namely building an accurate and robust model. This tool, once deployed can aid in the development of personalized therapies for these subtypes by aiding in clinical trials and also in the future, aid in application of personalized therapies and monitoring of patients, which means improving outcomes for the patients falling in the widely prevalent group 3 and group 4 subgroups by the application of better treatments and avoiding negative side effects of current treatments. Moreover the cross platform approach enables the application of the classification solution on a more recent version of DNA array, however the latest version (EPICv2) was not incorporated due to unavailability of data. However, while data build on the latest array solution probes classification ability on both HM450 and EPIC assayed samples (a notion of backwards compatibility), essentially covering a wider solution reach. It also explain the benefits of a cross platform solution to the research community.

Retrospectively, the solution has some limitation. Due to time constraint the classifier could not be deployed. Additionally, the solution maximally focused on avoiding any form of data leaks by segregating all train aspects from the test aspects, however due to computational and time constraints, NMF reduction and projection was not performed fold wise within the 10-Fold cross validation during hyper-parameter tuning resulting in a slight inflation in hyperparameter tuning cv score. Moreover, the model could not be validated on independent sets due to the unavailability of further data. It is also known that the DNA methylation array versions concerning this study have been deprecated, however, on the latest EPICv2 version data yet do not exist and hence a solution could not be made. Despite this the current solution even based on the deprecated versions is very useful as a lot of samples exist on the deprecated platforms and could benefit from this classifier. Lastly, survival analysis being crucial to medulloblastoma studies, could not be performed due to the unavailability of data because of ethical constraints. Considering AI ethics, measures were taken to conform to this. Sex related probes were removed to prevent gender bias. Ethnicity related bias was also considered, however, no actions or analysis was performed due to time constraints. No potential privacy concerns were noted as the model was not exposed to any patient identification data. In future's perspective, first the data leak in hyper-parameter tuning could be resolved which could lead to better selection of hyperparameters and hence also improve model's accuracy. Next, the model is planned to be deployed publicly through a web app enabling predictions and some analytics to open up the solution for users as the public availability of this classifier is the main end goal of this study given established need and impact. Additionally other medulloblastoma subgroups could also be integrated in this solution to build a one stop cross platform solution for all medulloblastoma subgroups. Similarly as soon as an large enough EPICv2 dataset is available, model adjustment should be performed to also incorporate the latest platform. Moreover, the area of ethnicity based bias can also be further explored to move towards a global classifier.

## VI. Conclusion

Concluding, this study resulted in building an accurate and reliable cross platform DNA methylation based ML classification solution for the purpose of classifying 8 novel medulloblastoma subtypes under group 3 and group 4, achieving an average weighted F1-score of 0.95 and average balanced accuracy of 0.957 across 3 test sets, with the ability to predict samples assayed on HM450 and EPIC arrays, The study demonstrated an ML solution approach using a standard ML pipeline starting with data preparation which included IDAT file processing to generate cross platform validation set, and then moving to preprocessing which included the use of NMF for dimension reduction and then moving to analytics, model building and evaluation. Although the tool could not be deployed due to time constraints, once deployed the tool would become the first publicly available classifier for these subtypes and could aid in the building and application of personalized treatments, and patient monitoring, essentially improving outcomes for a majority prevalence Group 3 and 4 population. The cross platform aspect enables the accommodation of a more recent version of methylation array while also being backward compatible to HM450, opening the benefits of the solution to a wider audience of samples considering both the platforms. While the solution is great in its current form, a set of limitations and future work are also noted. The most important amongst these is to resolve for the minor data leak happening in hyper-parameter tuning and deploying the model and analytics for public use by serving it through a webapp.

ACKNOWLEDGEMENTS

We would like to thank Dr. Ed Schwalbe (Northumbria University) and Professor Steve Clifford's lab (Newcastle University) for their valuable guidance and continued support throughout this project. Their contributions were instrumental in strengthening the translational relevance of this work.